\documentclass[12pt]{JHEP3}

\usepackage{graphicx}
\usepackage{verbatim}
\usepackage{latexsym}
\usepackage{amsmath,amsfonts,amssymb,amsthm,wasysym}
\def\bseq{\begin{subequation}}  
\def\eseq{\end{subequation}}
\def\bsea{\begin{subeqnarray}}  
\def\esea{\end{subeqnarray}}

\newcommand{\bbox}{\lower.2ex\hbox{$\Box$}}

\newcommand{\beq}{\begin{equation}}
\newcommand{\eeq}{\end{equation}}
\newcommand{\bea}{\begin{eqnarray}}
\newcommand{\eea}{\end{eqnarray}}
\newcommand{\ena}{\end{eqnarray}}

\newcommand {\non}{\nonumber}

\newcommand{\Tr}{{\rm Tr}}

\newcommand{\be}{\begin{equation}}
\newcommand{\ee}{\end{equation}}
\newcommand{\mc}{\mathcal}
\def\tr{{\rm tr}\ }

\title{\begin{center} Supersymmetric $D$-branes 
\\ on $SU(2)$ structure manifolds \end{center}}
\author{Alberto Mariotti \\ University of Milano-Bicocca and INFN Milano-Bicocca \\
Piazza della Scienza 3, 20126 Milano, Italy\\ 
\email{alberto.mariotti@mib.infn.it}}
\abstract{
We employ generalized complex geometry 
to investigate supersymmetric embeddings of $D$-brane 
probes 
in a large class of $SU(2)$ structure manifolds.
This class includes the gravity dual of mass deformation and marginal beta 
deformation of $\mathcal{N}=4$ SYM gauge theory.
We find supersymmetric configurations 
of 
$D$-branes 
with different dimensionality and propose their interpretation in the dual gauge theory. 
}

\begin{document}

\section{Introduction}
Strings and supergravity backgrounds with non trivial RR and NS fluxes 
are intensively studied
in the AdS/CFT correspondence \cite{MaldaWitten} 
and in string compactification (see \cite{Mariana} and reference therein), 
in order to
 find string models holographically dual to more realistic gauge theories or to obtain 
sensible phenomenology from compactification.
Here $D$-branes are successfully 
used as probes to explore the geometric properties of known backgrounds,
and to provide further insights in the gauge/gravity duality. 
We focus on type $IIB$ supergravity solutions
which
preserve four dimensional 
Poincar\'e invariance and $\mathcal{N}=1$ supersymmetry. 
They correspond to
a warped product of the four dimensional Minkowski spacetime 
and an internal six dimensional manifold 
$\mathcal{M}$, which can support fluxes.
In the presence of non trivial background fluxes, the back-reacted internal manifold $\mathcal{M}$ 
is no longer Calabi Yau. 
There are special classes of solutions \cite{GKP} 
where the internal manifold is conformal Calabi Yau, but
in general \cite{Grana1,Grana2} the internal manifold with fluxes can be 
far different from the 
Calabi Yau case.
 The formalism of $G$-structures \cite{Gauntlett} and Generalized Complex Geometry (GCG) 
 \cite{Hitchin,Gualtieri,Grana1,Grana2} provide powerful tools to describe such manifolds. 
In GCG the basic objects are pure spinors, formal sums of
even and odd forms. Their existence imposes topological constraints
on the tangent and cotangent bundles of the internal manifold. 
Supersymmetry requires that the internal manifold has a $SU(3)\times SU(3)$ 
structure on $T_M \oplus T_M^*$, which may be further restricted to 
$SU(3)$ or $SU(2)$ structures on $T_M$. The $SU(3)$ structure has been much studied, 
e.g.\cite{esseu3},
while the $SU(2)$ case has been explored in \cite{Dallagata} and, 
using GCG, in \cite{Zaffa}.
 As a matter of fact, supergravity solutions with fluxes dual to
 massive and marginal deformations of superconformal gauge theories
 are expected to be described by
 $SU(2)$ structure manifolds. Such manifolds are characterized 
 by the existence of a globally defined nowhere vanishing vector field. 
 
 In the GCG language the 
  preservation of $\mathcal{N}=1$ supersymmetry
  is achieved by imposing
  a pair of differential equations for the pure spinors. 
The authors of \cite{Zaffa} made an ansatz for pure spinors 
of $SU(2)$ structure manifolds
 and performed a detailed analysis of these pair of
 supersymmetry equations.
 Their ansatz 
  covers a large class of solutions.
  In particular the
 Pilch Warner \cite{PW} and the Lunin Maldacena \cite{LuninMalda} 
 ones are included: they are the gravity duals
 of the single mass deformation and of the beta marginal deformation
 of $\mathcal{N}=4$ SYM, respectively. 
 
In the GCG framework the supersymmetry conditions for $D$-branes probing
$SU(3)\times SU(3)$ backgrounds have been established in \cite{Koerber,Martucci} (see also
\cite{Martucci2}).
They are a set of 
constraints on the pull back of the pure spinors on the world
volume of the $D$-brane.
In \cite{Martucci} the supersymmetry conditions were given for 
$D$-branes filling Minkowski space time (space time filling), 
filling 
three space time directions
(domain walls)
 and 
  two space time directions
 (effective strings). 

The addition of $D$-brane probes to the class of solutions of \cite{Zaffa} can
provide other interesting tests of the $AdS/CFT$ correspondence.
Supersymmetric configurations of 
$D$-branes 
 can identify the moduli space
of vacua of the dual gauge theory, in both the abelian and the non abelian branches.
$D5$ domain wall like configurations can lead in the dual description 
to three dimensional defects,
interacting with the conformal four dimensional gauge theory; 
the defect gauge invariant operators can then be mapped into the Kaluza Klein modes
of the wrapped brane \cite{Ooguri}.
The addition of space time filling $D7$-branes corresponds to adding massless or massive 
flavours \cite{KarchKatz} and their 
fluctuations give the meson spectrum of the dual flavoured gauge theory.

In \cite{Zaffa} the space time filling $D3$-brane configurations 
have been analyzed and
it was shown that the supersymmetry conditions for such branes
 reproduce the mesonic moduli space of vacua of 
 the dual field theory. Moreover the $D5$-brane 
 configuration with world volume flux, related to the non abelian phase of
 the beta deformed gauge theory \cite{LuninMalda,Nonabe}, was recovered.\\

In this paper we 
investigate
new supersymmetric 
$D$-brane configurations in the class of $SU(2)$ structure manifolds 
of \cite{Zaffa}, and we propose the dual gauge theory interpretation
as well as possible applications of the results.

We look for supersymmetric $D5$ domain wall like configurations 
finding a supersymmetric embedding which can be
used to holographically study three dimensional  
 defects coupled to the massive deformation of $\mathcal{N}=4$ SYM. 
 
 We study a supersymmetric embedding
of space time filling $D5$-branes with non trivial world volume flux 
in the Pilch Warner solution.
 
We explore different $D7$ supersymmetric 
embeddings suitable for adding flavour to the whole class of solutions,
suggesting in each case the dual flavored gauge theory.
These embeddings identify supersymmetric four cycles.
Although the formalism we adopt does not apply to the non static case,
these supersymmetric four cycles should be mapped, with a strategy similar
to \cite{Mika}, to non static configurations of $D3$ branes (giant gravitons) in
this class of backgrounds\footnote{For giants in the beta deformed background see \cite{Marco}.} .

Finally, we find supersymmetric configurations of $D3$ and $D7$ branes which
behave as effective strings in the four dimensional gauge theory description.

The paper is organized as follows.
In section 2 we outline the spinor ansatz for $SU(2)$ structure manifolds \cite{Zaffa}
and 
in section 3 the GCG supersymmetry conditions for $D$-branes \cite{Martucci}.
In section 4, after a brief survey of the supersymmetric family of backgrounds
which includes the PW flow, 
we look for supersymmetric embeddings of $D$-branes.
We present different $D$-brane configurations 
and we solve their supersymmetry conditions,
identifying supersymmetric embeddings.
We give some details on the computations and
we interpret the supersymmetric configurations in
the dual gauge theory.
The same analysis is carried out for $D$-brane probes in the
LM geometry in section 5.
In the appendices we recall some useful definitions.

\section{$SU(2)$ structure manifolds and pure spinors}
The ten dimensional metric is 
\be
ds_{10}^2=e^{2 A} \eta_{\mu \nu} dx^{\mu} dx^{\nu}+ds_6^2
\ee
 where the warp factor $A$ is a function of the internal coordinates.
 The internal six dimensional manifold has $SU(2)$ structure.
An $SU(2)$ structure is characterized by two nowhere vanishing spinors which are never parallel
\be
\label{spinors0}
\eta_{+} \qquad \chi_{+}=\frac{1}{2} z \cdot \eta_{-}
\ee
where $\eta_{-}$ is the complex conjugate of $\eta_{+}$ and we denote with $\cdot$
the Clifford multiplication $z_{m} \gamma^{m}$.
The six dimensional chiral spinors $\eta^{i}_{\pm}$, which are the supersymmetry parameters,
 are
then constructed
 \be
\label{spinors1}
\eta_{+}^1=a \eta_+ + b \chi_+ \qquad
\eta_+^2=x \eta_+ + y \chi_+ 
\ee
with $a,b,x,y$ complex functions of the internal coordinates. 
The ten dimensional supersymmetry parameters can be written as 
\bea
\epsilon_1=\zeta_+ \otimes \eta_+^1+\zeta_- \otimes \eta_-^1 \\
\epsilon_2=\zeta_+ \otimes \eta_+^2+\zeta_- \otimes \eta_-^2
\eea 
where $\zeta_{\pm}$ are four dimensional chiral spinors.
Given the never vanishing spinors just introduced, 
a $SU(2)$ structure manifold admits the following 
globally defined 
forms built as bilinears in the spinors
\bea
&&j=\frac{i}{2} \chi_{+}^{\dagger} \gamma_{m n}\chi_+ dx^m \wedge d x^n-
\frac{i}{2} \eta_{+}^{\dagger} \gamma_{m n} \eta_+ dx^m \wedge d x^n \\
&& \omega=-i \chi_{+}^{\dagger} \gamma_{m n}\eta_+ dx^m \wedge d x^n \\
&& z=-2 \chi_{-}^{\dagger} \gamma_{m }\eta_+ dx^m  
\eea
where $z$ is a complex 1-form, $j$ a real 2-form, and $\omega$ a (2,0)-form satisfying
\be
\omega \wedge j=0 \qquad
 j \wedge j=\frac{1}{2} \omega \wedge \bar \omega \qquad
 z \llcorner j=z \llcorner \omega=0
\ee
The 1-form $z$ is the 
globally defined complex vector 
characterizing the $SU(2)$ structure.

In GCG the relevant equations can be written 
in terms of poliforms with definite parity, 
the pure spinors. They are bispinors built by
tensoring the supersymmetry parameters of the internal manifold
\bea
\Phi_1=\eta_+^1 \otimes \eta_+^{2 \dagger}\\
\Phi_2=\eta_+^1 \otimes \eta_-^{2 \dagger}
\eea
and are annihilated by six combinations of Clifford(6,6) gamma matrices.
From (\ref{spinors1}) they read
\bea
\label{pure}
\Phi_1&=&\frac{1}{8}[a \bar x e^{-i j}+b \bar y e^{i j}-i (a \bar y \omega+ \bar x b \bar \omega)]
\wedge e^{z \wedge \bar z /2}\\
\Phi_2&=&\frac{1}{8}[i(b y  \bar \omega -a x \omega)+ (b x  e^{i j}- a y e ^{-i j})] \wedge z \non
\eea
The $SU(3)$ structure case is for $b=0=y$. 

The ansatz used in \cite{Zaffa} for the six dimensional supersymmetry parameters is the following
\be
\label{ansatz1}
\eta_+^1= a \eta_+ + b \chi_+   \qquad
\eta_+^2= -i(a \eta_+ - b \chi_+) 
\ee
where the functions of (\ref{spinors1}) are parametrized  
as
\be
\label{ansatz2}
a=i x = i e^{A/2} \cos \phi   \,  e^{i \alpha}\qquad
b=-i y=-i e^{A/2} \sin \phi \, e^{i \beta}
\ee
Here $\cos \phi$, $\sin \phi$, $\alpha$ and $\beta$ are functions of the internal coordinates.
The two supersymmetry parameters $\eta_+^1, \eta_+^2$ can be brought to the form (\ref{ansatz1})
if and only if $\text{Re}(a \bar x+ b \bar y)=0$. 
This corresponds to admit a non trivial
mesonic moduli space of vacua \cite{Zaffa}.

We are interested in $D$-branes probing
 the class of backgrounds specified by the ansatz 
(\ref{ansatz1}), (\ref{ansatz2}). This contains
a family of supersymmetric backgrounds with constant dilaton 
(which itself includes the PW flow), and the gravity dual of
beta deformation.
Since the norms of the spinors $\eta_{1}$ and $\eta_2$ are equal,
 supersymmetric $D$-branes are admitted \cite{Martucci}.

\section{Supersymmetry conditions for probe $D$-branes}
In GCG the main tool to analyze supersymmetric embeddings of
$D$-branes is the generalized calibration
introduced in \cite{Koerber,Martucci}.
We will 
consider space time filling branes (STF), domain walls (DW)
and effective strings (ES) wrapping
a submanifold $\Sigma$ of the internal manifold. 
The supersymmetry conditions for these extended objects
in terms of the pure spinors and their projection on the world 
volume read\footnote{We do not consider the orientation conditions on these objects.}
\bea
\label{susyeq1}
P_{\Sigma}[\text{Im} (i e^{i \theta} \Phi_a)] \wedge e^{\mc{F}} &=&0\\
\label{susyeq2}
P_{\Sigma}[(i_n+g_{nm} dx^m \wedge)\Phi_b] \wedge e^{\mc{F}} &=&0 \qquad a,b=1,2
\eea
where $g_{nm}$ is the internal metric, $i_{n}$ and $dx^m \wedge$ are the usual
operators mapping a $p$ form in a $p-1$ and $p+1$ form respectively, and 
finally \footnote{We are using the conventions of \cite{Grana1,Grana2,Zaffa} which differs for 
an $H_{NS}$ sign with \cite{Martucci}.}
$\mathcal{F}=F-P_{\Sigma}[B]$, where $F$ is the world volume flux. 
The pullback on the world volume of the $D$-brane is denoted by $P_{\Sigma}$.
Space time filling branes, domain walls and effective strings are 
summarized in Table 1,
where $\theta_{DW}$ is an arbitrary constant \cite{Martucci}.  
\begin{center}
\begin{tabular}{c| c| c| c|}
& $\theta$ & a & b \\
\hline
STF & 0 & 1& 2 \\
\hline
DW & $\theta_{DW}$ &2&1\\
\hline 
ES & $-\frac{\pi}{2}$ &1&2\\
\hline
\end{tabular} 
\\
~~\\
\bf{Table 1}
\end{center}

The same dictionary of
\cite{Martucci} is used to label the possible embeddings. However,
since the internal manifold is non compact, we should 
distinguish between the cases
when the wrapped submanifold $\Sigma$
is itself compact or 
non compact.
We will comment on this point where needed.

\section{$D$-branes on the family of supersymmetric backgrounds}
\subsection{The family of supersymmetric backgrounds} \label{supclass}
We now  briefly review 
the family of supersymmetric backgrounds analyzed in \cite{Zaffa}
which includes the PW flow \cite{PW}. 
The PW solution is the gravity dual of the massive deformation of $\mathcal{N}=4$ SYM
\be
\label{massdef}
W=h \Tr \Phi_3[ \Phi_1,\Phi_2]+m \Tr \Phi_3^2
\ee
which flows in the IR to a non trivial fixed point \cite{LS}.
The gravity dual is asymptotically $AdS$ in the UV and warped $AdS$ in the IR.
It is included in the following more general ansatz \cite{Zaffa}
which is a family of supersymmetric backgrounds  
\be
\label{6dmet}
ds^2_6=e^{-2 A} \left( \eta_i A_{i \bar j} \bar \eta_{\bar j} + z \bar z \right) 
\qquad i,j=1,2
\ee
where $z$ is the 
globally defined vector characterizing the $SU(2)$ structure.
The matrix $A_{i \bar j}$ is hermitian, 
and the vielbeins are defined in terms of local complex coordinates $z_i$
\bea
&& z_1=\rho_1+i \sigma_1  \qquad z_2= \rho_2+i \sigma_2 \qquad z_3= \log u +i \sigma_3 \\
&&\eta_1=dz_1+\alpha_1 dz_3 \qquad \eta_2=dz_2+\alpha_2 dz_3 \qquad z=\sqrt{a_3} u d z_3
\eea
with $a_3$ real and  $\alpha_i$ complex functions of $z_i$.
The globally defined 
two forms are
\bea
\label{j2form}
&& j= \frac{i}{2} A_{i \bar j} \, \eta_i \wedge \eta_{\bar j} \\
&& \omega=i \sqrt{\det A} \, \eta_1 \wedge \eta_2
\eea
There are also non trivial RR and NS fluxes 
\bea
&&*F_5=- e^{-4 A} d(e^{4A} \cos 2 \phi )\\
&& C_2=\text{Re}[\frac{2 i e^{i(\alpha-\beta)} \sqrt{\det A}}{e^{2A} \sin 2 \phi} (dz_1\wedge dz_2-\sin^2 \phi \, \eta_1 \wedge \eta_2)]\\
\label{Bfield}
&& B_2=-\text{Im}[\frac{2 i e^{ i(\alpha-\beta)} \sqrt{\det A}}{e^{2A} \sin 2 \phi} (dz_1\wedge dz_2-\sin^2 \phi\, \eta_1 \wedge \eta_2)]
\eea
The dilaton is constant, parametrising the RG line of dual conformal
gauge theories.

The supersymmetry equations for this background \cite{Zaffa} imply that
$\alpha=\frac{1}{2}(\sigma_1+\sigma_2+3 \sigma_3)$, 
$\beta=-\frac{1}{2}(\sigma_1+\sigma_2-\sigma_3)$ and that
the functions $a_3, \alpha_i, A_{i \bar j}$ can be obtained as derivatives
of a single function 
$F(z_i,\bar z_{\bar j})$. 
These are all real for the subclass of this family of backgrounds which have an 
$U(1)^3$ symmetry, i.e. when the function $F(z_i, \bar z_{\bar j})$ does not
depend on the phases $\sigma_i$. 
We call this the \emph{toric subclass};
the PW flow belongs to it. 

The detailed expressions for the family of backgrounds and how to recover the PW flow
are reported in the Appendix A.

The pure spinors (\ref{pure}) are constructed
with the rescaled forms $z \to e^{-A} z$ and $(j,\omega)$ $\to$ $(e^{-2 A} j, e^{-2 A} \omega)$ 
which refer to the complete six dimensional metric   (\ref{6dmet}).\\

We look for supersymmetric embeddings of
$Dp$-branes (with world volume coordinates $\xi_a$ ($a=0,\dots, p$))
in this family of supersymmetric backgrounds,
allowing in one
case for non trivial world volume gauge flux.
The main tools are the conditions  (\ref{susyeq1}),(\ref{susyeq2}).

Even if the family of backgrounds is larger,
we shall take the PW solution as a paradigm
for the gauge theory dual interpretation of
the brane configurations.

\subsection{$D5$ domain walls}
We study now a supersymmetric $D$-brane probe
placed at  $x_3=0$ and which fills three space time dimensions
$(\xi_0,\xi_1,\xi_2)=(x_0,x_1,x_2)$. 
 It can be viewed
as a domain wall solution separating supersymmetric vacua.
However, when the wrapped cycle is non compact, the domain wall interpretation
would imply an infinite potential barrier. 
Instead in the AdS/CFT interpretation it is a three dimensional defect 
coupled to the
four dimensional dual gauge theory.

In the $AdS_5 \times S^5$ case there are non trivial supersymmetric  embeddings where
a $D5$-brane wraps an $AdS_4$ inside the $AdS_5$ plus 
a trivial 2-sphere inside the $S^5$ \cite{Randall}. 
The $D5$ brane should shrink around this 2-sphere but the correspondent 
tachionic mode does not lead to instability because its mass is above the BF bound \cite{BF}. 
This configuration has been studied in \cite{Ooguri}
as a three dimensional defect in 
$\mathcal{N}=4$ SYM.
 
We look for 
similar
configurations of
$D5$-brane 
in the family of supersymmetric backgrounds of section \ref{supclass}.
We attempt the following three cycle embedding 
\be
\label{d5dw1}
z_k=e^{i \tau_k} ( \xi_{k+2}+i c_k) 
\qquad \bar z_k=e^{-i \tau_k} (\xi_{k+2}-i c_k)  \qquad k=1,\dots,3
\ee
with $\tau_k$ and $c_{k}$ constants, and with no world volume flux, $F=0$.
This ansatz covers for example the real slice ($\tau_k=0, \, \forall k$)
and the imaginary slice ($\tau_k=\frac{\pi}{2}, \, \forall k$).

We restrict ourselves to the \emph{toric subclass}.
The complex functions $\alpha_i, A_{i \bar j}$ 
characterizing the metric are then real and the computations simplify.
We compute the supersymmetry conditions  
(\ref{susyeq1}) and (\ref{susyeq2}) in the DW case of Table 1.

The supersymmetry condition  (\ref{susyeq1}) results
\be
\label{d5dwcond}
P_{\Sigma}[\text{Im} (i e^{i \theta_{DW}} \Phi_2)] \wedge e^{\mathcal{F}}=
\frac{1}{8} \text{Im} [e^{-2 A}\sqrt{a_3} u \sqrt{\det A} \, 
e^{i (\theta_{DW}+2 \beta-\tau_1-\tau_2+\tau_3)}   ]
\, d\xi_3 \wedge d\xi_4 \wedge d\xi_5
\ee
where the functions are intended evaluated on the world volume.
A choice of the constant phase $\theta_{DW}$ can make it vanish 
only if the phase factor $\beta$ does not depend on the embedding coordinates $\xi_{k+2}$.
This can be achieved taking the real slice ($\tau_k=0, \, \forall k$), such that
$\beta=-\frac{1}{2}( c_1+c_2-c_3)$.
Then we choose  $\theta_{DW}=-2 \beta$ and the expression
(\ref{d5dwcond}) vanishes.

For the real slice ($\tau_k=0, \, \forall k$), a detailed analysis shows that the
 supersymmetry conditions  (\ref{susyeq2}) are
 satisfied provided $\alpha=\beta+\frac{\pi}{2}$. 
This implies the following relation between the constants $c_k$
\be
\label{conddw3}
c_1+c_2+c_3=\frac{\pi}{2}
\ee

Hence we conclude that for the \emph{toric subclass} 
a $D5$ brane embedded as in 
(\ref{d5dw1}) with $\tau_k=0$, with the constants $c_k$ satisfying (\ref{conddw3})
and with $\theta_{DW}=( c_1+c_2-c_3)$ is supersymmetric.
In particular, such $D5$ brane 
is supersymmetric in the PW flow, since it belongs to the 
\emph{toric subclass}. 
In the PW geometry (see the appendix A)
the $D5$ brane
fills the three radial directions.

This embedding 
can be used
to study three dimensional
 defects in the massive deformation of $\mathcal{N}=4$.
The $c_i$ 
give the distance between the supersymmetric $D5$-brane and the $D$-branes which 
generate the background. 
They represent masses for the 3D hypermultiplet of the defect theory.

\subsection{Spacetime filling $D$-branes}
In this section we study $D$-brane probes filling all the Minkowski directions 
$\xi_{\mu}=x_{\mu}$ ($\mu=0,\dots,3$).
The supersymmetry conditions are (\ref{susyeq1}) and (\ref{susyeq2}) in the
STF case of Table 1.
We analyze here supersymmetric
$D5$-brane embeddings with world volume flux,
 and $D7$ flavour branes.

\subsubsection{$D5$-branes}
We take the following two cycle embedding $\Sigma$ for a $D5$ brane
probing the background of section \ref{supclass}
\be
\label{stfd50}
z_k=e^{i \tau_k} (\xi_{k+3}+i c_k) \qquad k=1,2 \qquad z_3=c_3+ i c_4
\ee
with $c_k$ and $\tau_k$ real constants. We allow for a generic world volume flux $F$.
The only non trivial supersymmetry conditions for this configuration are the
(\ref{susyeq1}) and the $z$ component of (\ref{susyeq2}), since $\Phi_2=\dots \wedge z$
and $P_{\Sigma}[z]=0$ from (\ref{stfd50}). 
The first one reads
\be
\label{d5stfeq1}
P_{\Sigma}[\text{Im} (i  \Phi_1)] \wedge e^{\mc{F}}=-
\frac{i e^{-A }}{16} (A_{1 \bar 2} e^{i (\tau_1-\tau_2)}-A_{2 \bar 1} e^{-i (\tau_1-\tau_2)}) \, d\xi_{4} \wedge d\xi_{5}
\ee
and does not depend on the two form flux $\mathcal{F}=F-P[B]$
 since $P_{\Sigma}[\text{Im} (i  \Phi_1)] |_0=0$.
This expression cannot be made vanishing in general by a simple choice of
the phases $\tau_1,\tau_2$. 
However, if we restrict ourselves 
to the \emph{toric subclass} the matrix $A_{i \bar j}$ is real and symmetric,
and $A_{1 \bar 2}=A_{2 \bar 1}$. 
If we then choose $\tau_1=\tau_2$ the expression (\ref{d5stfeq1}) vanishes.

We compute the $z$ component of the second supersymmetry condition
\be
\label{eq1d5stf}
P_{\Sigma}[(i_{z}+g_{z \bar z} \bar z \wedge) \Phi_2] \wedge e^{\mc{F}}=
-\frac{i e^{- 2A}}{8} ( F_{\xi_4 \xi_5} e^{2 A}  e^{i (\alpha+\beta)} \sin 2 \phi + 
\sqrt{\det A} e^{-i(\tau_1+\tau_2-2 \beta)} ) \, d\xi_{4} \wedge d\xi_{5}
\ee
where 
$F_{\xi_4 \xi_5}$ is the world volume flux. 
The expression (\ref{eq1d5stf}) vanishes if we turn on 
\be
\label{d5stfflux}
F=-e^{-i (\tau_1+\tau_2+\alpha-\beta)} \frac{\sqrt{\det A}}{e^{2 A}\sin 2 \phi}  d\xi_4 \wedge d\xi_5
\ee
which for consistency 
should be real.
The choices
\be
\label{fixd5}
\tau_1=\tau_2=0 \qquad \alpha-\beta=c_1+c_2+c_3=0
\ee
make the flux (\ref{d5stfflux}) real, since the phase factor in 
(\ref{d5stfflux}) is now independent of the embedding coordinates $\xi_{k+3}$
and moreover it vanishes.
We conclude that the choices 
(\ref{d5stfflux}) and (\ref{fixd5}) make the $D5$ brane configuration 
(\ref{stfd50}) supersymmetric in the \emph{toric subclass}.

However particular
care is needed in considering this embedding; indeed we 
 observe that the $D5$ brane 
 wraps a non compact submanifold and then
 the flux $F$ is along non compact coordinates
 (see for example the coordinates for the PW geometry in appendix A).

\subsubsection{$D7$ flavour branes}
Here we look for supersymmetric $D7$-brane embeddings suitable
for adding flavours to the family of 
backgrounds
of section \ref{supclass}. The $D7$ branes should wrap a 
non compact four cycle in order to make the flavour
symmetry group global. 
Adding $N_f$ $D7$ branes 
on this non compact
four cycle is dual to add $N_f$ flavours with
symmetry group $SU(N_f)$ to the $SU(N_c)$ gauge theory provided $N_f < N_c$, so
that the back-reaction of the $D7$-branes can be neglected.
The shape of the $D7$ supersymmetric embedding sets 
the interaction terms in the superpotential between the flavours and the chiral superfields of
the dual gauge theory as well as possible masses for the flavours.  

In a $SU(2)$ structure manifold the globally defined vector $z$ naturally
identifies a four dimensional submanifold $\Sigma$ where $P_{\Sigma}[z]=0$.
Thus we attempt 
the embedding with $P_{\Sigma}[z]=0$, i.e. 
we place $D7$ branes as 
\bea
&&x_{\mu}=\xi_{\mu} \qquad \mu=0,\dots,3 \non \\
\label{embd7}
&& z_k=\xi_{k+3}+i \xi_{k+5} \qquad k=1,2 \qquad z_3=\log m_0
\eea
with no world volume flux, $F=0$, and where $m_0$ is an arbitrary constant.
The first supersymmetry condition (\ref{susyeq1}) can be analyzed 
by keeping the $4,2,0$ forms of the pulled back
pure spinor $\Phi_1$
\bea
&& i \Phi_1 |_{0}=-\frac{e^{A}}{8}(\cos^2 \phi-\sin^2 \phi)  \non \\
&& i  \Phi_1 |_{2}=\frac{i e^{-A}}{8}(j+
\cos \phi \sin \phi (e^{i(\alpha-\beta)} \omega - e^{-i(\alpha-\beta)}\bar \omega)) \non \\
&& i \Phi_1 |_{4}=\frac{e^{-3 A}}{16} (\cos^2 \phi-\sin^2 \phi) j \wedge j \non 
\eea
Taking the imaginary part of these expressions we obtain 
\be
\label{eqd7d}
P_{\Sigma}[\text{Im} (i \Phi_1)] \wedge e^{-P[B]} =-\frac{e^{A}}{8} P[j] \wedge P[B]=0
\ee
This vanishes given the explicit expressions of $j$ (\ref{j2form}) and $B$ (\ref{Bfield}) 
and reminding $P_{\Sigma}[z]=0$.
The only non trivial supersymmetry condition of (\ref{susyeq2}) is
on the $z$ component. The projection on the pure spinor $\Phi_2$ is
\bea
P_{\Sigma}[(i_{z}+g_{z \bar z} \bar z \wedge) \Phi_2] &=&
\frac{1}{8}(
- i e^{i(\alpha+\beta)}\sin 2 \phi +
e^{-2 A} e^{2 i \alpha} \cos^2 \phi \, \omega+\non \\
&&
+e^{-2 A} e^{2 i \beta}\sin^2 \phi \, \bar \omega +
\frac{i}{2} e^{-4A}  e^{i (\alpha+\beta)} \sin 2 \phi \, j \wedge j) \non 
\eea
The pullback of the NS two form (\ref{Bfield}) is
\bea
P_{\Sigma}[B]&=&-\frac{\sqrt{\det A} \cos^2 \phi}{e^{2 A} \sin 2 \phi} 
\big( \, e^{i (\alpha-\beta)} (d\xi_4+i d\xi_6)\wedge(d\xi_5+i d\xi_7)+\\ 
&&~~~~~~~~~~~~~~~~~~~~~
~~+e^{-i (\alpha-\beta)} (d\xi_4-i d\xi_6)\wedge(d\xi_5-i d\xi_7) \, \big) \non
\eea
We then compute the terms which
contribute to the $z$ component of 
(\ref{susyeq2})\footnote{
We denote the volume on the wrapped cycle with $d\text{Vol}_{\Sigma}=(-4 d \xi_4 \wedge d \xi_5 \wedge d \xi_6\wedge d \xi_7)$.
}
\bea
&&P_{\Sigma}[(i_{z}+g_{z \bar z} \bar z \wedge) \Phi_2] |_4=
\frac{i e^{i(\alpha+\beta)}}{e^{4A}16} \det A \cos \phi \sin \phi  \, \,
d\text{Vol}_{\Sigma} \non \\
&&P_{\Sigma}[(i_{z}+g_{z \bar z} \bar z \wedge) \Phi_2] |_2 \wedge (-P_{\Sigma}[B])=
\frac{i e^{i(\alpha+\beta)}}{16} \frac{\cos \phi \det A}{e^{4 A}\sin \phi}(\cos^2 \phi -\sin^2 \phi)  \, 
d\text{Vol}_{\Sigma} 
\non
\\
&&P_{\Sigma}[(i_{z}+g_{z \bar z} \bar z \wedge) \Phi_2] |_0 \wedge \frac{1}{2} P_{\Sigma}[B] \wedge P_{\Sigma}[B]=
-\frac{i e^{i(\alpha+\beta)}}{16} \frac{ \cos^3 \phi \det A}{e^{4 A }\sin \phi}  \,
d\text{Vol}_{\Sigma}
\non
\eea
Adding these three contributions we conclude that
\be
P_{\Sigma}[(i_{z}+g_{z \bar z} \bar z \wedge) \Phi_2] \wedge e^{-P[B]}=0
\ee
Then the configuration (\ref{embd7}) is supersymmetric for
the whole family of backgrounds considered in section \ref{supclass},
not only the \emph{toric subclass}.

\paragraph{Other flavour embeddings}
We look also for other $D7$ brane embeddings which preserve supersymmetry
in the supersymmetric family of backgrounds of sec \ref{supclass}.
The computations of the 
supersymmetry conditions (\ref{susyeq1}) and (\ref{susyeq2}) 
are less easy but can be done with the same
procedure outlined above. 
We list the relevant results.

We can place the $D7$ brane orthogonal to
one of the other complex coordinates
\be
\label{embd71}
z_k=\log m_0 \qquad z_{j}=\xi_{4}+i \xi_{5} \qquad z_3=\xi_{6}+i\xi_7 \qquad k \neq j=1,2
\ee
and after a long computation we find that this is a
supersymmetric configuration, satisfying (\ref{susyeq1}) and (\ref{susyeq2}). 

Other possible embeddings 
are submanifolds like
the one suggested in \cite{KarchKatz},
with chiral symmetry breaking. 
We observe that
the complex coordinates we are using (see the Appendix A)
are
the exponential of the usual complex coordinates
which are in correspondence with the chiral adjoint fields. 
Hence we consider embeddings 
like $e^{z_i} e^{z_j}=m_0^2$.
We have to distinguish between
two different cases. The first one involves the $z_3$ component 
\bea
&&e^{z_j} e^{z_3}=m_0^2 \non \\
&&z_k=\xi_4+i \xi_5  \qquad  z_j=\xi_6+i \xi_7 \qquad
 z_3=\log m_0^2-(\xi_6+i \xi_7)  \qquad k \neq j=1,2 \non
\eea
This configuration turns out to be non supersymmetric.

The second case does not involve the $z_3$ coordinate
\bea
&&e^{z_1} e^{z_2} = m_0^2 \non \\
\label{embd72}
&&z_1=\xi_4+i \xi_5 \qquad  z_2=\log m_0^2 - (\xi_4+i \xi_5) \qquad z_3=\xi_6+i \xi_7
\eea
and it results supersymmetric.

\paragraph{The dual flavoured gauge theory}
The $D7$
supersymmetric embeddings presented here (\ref{embd7}), (\ref{embd71}), (\ref{embd72})
can be 
used to add flavours to the PW flow.

If we add $N_f$ $D7$-branes in the configuration
(\ref{embd7}) 
the dual gauge theory is $\mathcal{N}=1$ SYM
with three chiral adjoint fields 
and $N_f$ massive flavours with mass $m_0$, with superpotential
\be
W=W_{\mathcal{N}=4}+m \Tr \Phi_3^2+ \tr Q \Phi_3 \tilde Q + m_0 \, \tr Q \tilde Q
\ee
where the first two terms are the mass deformation of 
$\mathcal{N}=4$ SYM (\ref{massdef}).
Since we are 
neglecting the back-reaction of the $D7$ branes,
the geometry filled by the $D7$-branes in the IR is warped $AdS_5$ and 
the theory flows to the same IR fixed point.
For
$m_0 \neq 0$, the $D7$-branes end before reaching the IR.

If we add $N_f$ $D7$-branes as in
(\ref{embd71}) the gauge theory dual 
is again
 $\mathcal{N}=1$ SYM with three chiral adjoint fields 
and $N_f$ massive flavours, with superpotential 
\be
W=W_{\mathcal{N}=4}+m \Tr \Phi_3^2+\tr \, Q \Phi_k \tilde Q+m_0 \, \tr Q \tilde Q \qquad k=1,2
\ee
The flavours $Q \tilde Q$ 
now couple to 
the massless adjoint field $\Phi_k$.

Finally,
if we add $N_f$ $D7$-branes embedded as 
(\ref{embd72})
the 
dual flavoured gauge theory is $\mathcal{N}=1$ SYM with three chiral adjoint fields 
and two different $N_f$ massive flavours, with superpotential
\be
W=W_{\mathcal{N}=4}+m \Tr \Phi_3^2+\tr \,  Q_1 \Phi_1 \tilde Q_1 + \tr \, Q_2\Phi_2 \tilde Q_2 +  
m_0 \, \tr (Q_1 \tilde Q_2+Q_2 \tilde Q_1)
\ee
where $Q_1$ and $Q_2$ denote the two flavours. This configuration
can be interpreted as two sets of $N_f$ $D7$-branes at $e^{z_1}=m_0$ and $e^{z_2}=m_0$ respectively,
 each supporting different flavours,
which are joint smoothly into one set of $N_f$ $D7$ branes wrapped on $e^{z_1} e^{z_2}=m_0^2$ 
\cite{KarchKatz}. On the dual gauge theory picture there are two flavour groups
$SU(N_f)_1 \times SU(N_f)_2$ broken to the diagonal subgroup by the mass term $m_0$.

\subsection{Effective Strings}
We take $D$-branes that fill two coordinates in the Minkowski space time,
for example at $x_2=x_3=0$, filling $\xi_0=x_0, \xi_1=x_1$.
They can be viewed as propagating strings in the four dimensional description.
However, when the wrapped cycle of the internal manifold is non compact, 
the effective string tension
in the four dimensional picture diverges. 
The supersymmetry conditions are the pair (\ref{susyeq1}) and (\ref{susyeq2}) in the
ES case of Table 1.
We find supersymmetric embeddings of both $D3$ and $D7$ branes
which involve non compact cycles in the internal manifold. 
 The $D3$ brane
 wraps a two cycle, whereas the $D7$ brane fills
the whole internal manifold. 
Our analysis concern the whole family of backgrounds presented in
 section \ref{supclass}.

\paragraph{$D3$ effective strings}
We place $D3$-brane probes filling two directions in the internal space. 
We fix 
the $z_3$ coordinate, i.e. $z_3=c_3 e^{i \tau_3}$ and we look for supersymmetric embeddings
filling $z_1$ and $z_2$. The 
embedding along the two complex coordinates, $z_k=e^{i \tau_{k}}(\xi_{k+1}+i c_k)$ for $k=1,2$ 
results non supersymmetric.

On the other hand, the non compact embedding where
we identify $z_1$ and $z_2$ except for constant phases and
shifts
\be
z_1=e^{i \tau_1} (\xi_2+c_1+i (\xi_3+c_2)) \qquad z_2=e^{i \tau_2} (\xi_2-c_1+i (\xi_3-c_2)) \qquad
z_3=c_3 e^{i \tau_3}
\ee 
results supersymmetric for any choice of the phases $\tau_k$ and of the real 
constants $c_k$.

\paragraph{$D7$ effective strings}
We probe the geometry with $D7$-brane covering the whole internal space
\be
z_k=\xi_{k+1}+i \xi_{k+4} \qquad k=1,\dots,3
\ee
By a long but straightforward computation we find that this is a supersymmetric embedding,
which satisfies the supersymmetry conditions.

\section{$D$-branes on the beta deformed background}
\subsection{Beta deformation of $\mathcal{N}=4$ SYM and its gravity dual}
The $\mathcal{N}=1$
beta deformed gauge theory is a marginal deformation \cite{LS} 
of the $\mathcal{N}=4$ SYM, with superpotential  
\be 
W_{\beta}=h \, Tr( e^{ i  \pi \beta} \Phi_1\Phi_2 \Phi_3 - e^{- i  \pi \beta}
 \Phi_1\Phi_3 \Phi_2  ) 
\ee
where $\Phi_i$ are the three chiral adjoint superfields, and $\beta$ 
a complex constant.  
We consider $\beta$ to be real; in this case it is usually denoted as
$\gamma$.
Besides the $U(1)_R$ symmetry, this theory has two global symmetries
$U(1)_a \times U(1)_b$ with charges 
\begin{center}
\begin{tabular}{c|c|c|c}
&$\Phi_1$&$\Phi_2$&$\Phi_3$\\
\hline
$U(1)_a$&0&1&-1\\
\hline
$U(1)_b$&-1&1&0\\
\end{tabular}
\end{center}
These two global symmetries 
were crucial in the
generating solutions technique of \cite{LuninMalda},
where the supergravity background
dual to such gauge theory has been obtained.
This background
 has been analyzed using generalized complex geometry in \cite{Zaffa}. 
The ten dimensional metric is
\be
\label{d6beta}
ds^2= e^{2 A} ds_{Mink}^2+ds^2_6,\qquad  ds^2_6=e^{-2 A} d\tilde s_6^2
\ee
where $\tilde ds^2_6$ is the rescaled internal metric.
The internal $SU(2)$ structure manifold
can be described by local complex coordinates
\bea
&&z_1=r \mu_1 e^{i \sigma_1}=r \cos \alpha e^{i(\psi-\varphi_2)} \non \\
\label{complexLM}
&&z_2=r \mu_2 e^{i \sigma_2}=r \sin \alpha \cos \theta e^{i(\psi+\varphi_1+\varphi_2)}  \\
&&z_3=r \mu_3 e^{i \sigma_3}=r \sin \alpha \sin \theta e^{i(\psi-\varphi_1)} \non
\eea
The almost complex structure can be expressed  \cite{Zaffa} 
in terms of 1-forms (for details see the Appendix B) which give the rescaled metric
a simple expression  
\be
\label{metricazaf}
d\tilde s_6^2=x_1^2+x_2^2+G(y_1^2+y_2^2)+ z \bar z
\ee
where 
\be
\label{zetc}
G=\frac{1}{1+\gamma^2 g} \qquad z=\frac{d(z_1 z_2 z_3)}{r^2 \sqrt{g}}  \qquad g=\mu_1^2 \mu_2^2+\mu_2^2 \mu_3^2+\mu_3^2 \mu_1^2 \qquad e^{2 A}=r^2
\ee
The background has non trivial dilaton, RR and NS fluxes 
\bea
&& e^{\phi}=\sqrt{G}\\
&& B_2=\gamma \sqrt{g} G \frac{y_1 \wedge y_2}{r^2}\\
&& F_3=12 \gamma \cos \alpha \sin^3 \alpha \sin \theta \cos \theta d\psi \wedge d\alpha \wedge d\theta \\
&& F_5=4 (\text{vol}_{AdS_5}+*\text{vol}_{AdS_5})
\eea
This solution differs from the family of backgrounds reviewed in
section \ref{supclass}, for example the dilaton is not constant here.
However it is an $SU(2)$ structure manifold which can be described by the
ansatz (\ref{ansatz1}) and (\ref{ansatz2}) for the spinors \cite{Zaffa}. 
The 1-form $z$ in (\ref{metricazaf})
is 
a globally defined vector.
The 2-forms $j$ and $\omega$ are
\bea
j&=& \sqrt{G} (x_1 \wedge y_1+x_2 \wedge y_2) \\
\omega&=& i (x_1 + i \sqrt{G} y_1)\wedge(x_2+i  \sqrt{G} y_2)
\eea
and 
\bea
&&a=i x = i e^{A/2} \cos \phi= \frac{i}{\sqrt{2}} e^{A/2} (1+\sqrt{G})^{\frac{1}{2}} \\
&&b=-i y =-i e^{A/2} \sin \phi=\frac{i}{\sqrt{2}} e^{A/2} (1-\sqrt{G})^{\frac{1}{2}}
\eea
The phases $\alpha$ and $\beta$ in (\ref{ansatz2})
are vanishing, $\alpha=\beta=0$.
Once again the pure spinors (\ref{pure}) are constructed
with the rescaled forms $(j,\omega) \to (e^{-2 A} j, e^{-2 A} \omega)$ and $z \to e^{-A} z$
which refer to the complete six dimensional metric  (\ref{d6beta}).
\\

We look for supersymmetric embeddings of $D$-branes in this
background employing the conditions (\ref{susyeq1}) and (\ref{susyeq2}).

\subsection{$D5$ domain walls} 
We look for $D5$-brane embeddings filling three directions in the internal manifold
and placed in Minkowski at  $x_3=0$ with
($\xi_{\mu}=x_{\mu}$, $\mu=0,1,2$).
We choose the following ansatz, which is supersymmetric 
 in the undeformed $\gamma=0$ case 
($AdS_5 \times S^5$),
\be
\label{d5dwlm}
z_k=e^{-i \tau_k} (\xi_{k+2}+i c_k) \qquad \bar z_k=e^{i \tau_{k}} (\xi_{k+2}-i c_k) \qquad k=1,\dots,3
\ee
where $\tau_k, c_k$ are arbitrary real constants. Computing the supersymmetry 
conditions (\ref{susyeq1}) and (\ref{susyeq2})\footnote{In the DW case of Table 1.}
 this embedding
results non supersymmetric for any choice of the constants $\tau_k, c_k$.
For instance in the simple case ($\tau_k=0,c_k=0$) the $z$ and
$\bar z$ components of the supersymmetry conditions (\ref{susyeq2}) can
be computed
\be
\label{dwLM}
\frac{1}{3}P_{\Sigma}[(g^{\bar z z} i_{z}+ \bar z \wedge) \Phi_2] \wedge e^{-P[B]}=
P_{\Sigma}[(g^{z \bar z} i_{\bar z}+ z \wedge) \Phi_2] \wedge e^{-P[B]}=
-\frac{i}{16} e^{- A} \gamma \sqrt{g \, G}
\ee
where the functions ($A$, $g$, $G$) are intended evaluated on the world volume.
The result (\ref{dwLM}) cannot vanish unless $\gamma=0$, i.e. the
undeformed case; hence the embedding
(\ref{d5dwlm}) is not supersymmetric in the beta deformed background.

\subsection{$D7$ flavour branes}
We look for supersymmetric  $D7$ configurations filling the Minkowski
space time
$\xi_{\mu}=x_{\mu}$ ($\mu=0,\dots,3$) and
 wrapped
on a non compact four cycle in the internal manifold,
suitable for adding flavour to the
beta deformed theory.
As already observed, an $SU(2)$ structure manifold is characterized by 
a globally defined vector ($z$), and a natural four cycle $\Sigma$ 
is 
where $P_{\Sigma}[z]=0$.
In the beta deformed background the vector z is (\ref{zetc}), and
the condition $P_{\Sigma}[z]=0$ implies, in complex coordinates, 
\be
\label{emb}
z_1 z_2 z_3=m^3
\ee
with $m$ constant. 

We then take the following four cycle embedding
for $D7$-branes
\be
\label{embd7lm}
z_k=\xi_{k+3} e^{i \xi_{k+5}} \qquad k=1,2  \qquad \qquad 
z_3=\frac{m^3}{\xi_4 e^{i \xi_{6}} \xi_5 e^{i \xi_7}} 
\ee
with no world volume flux, i.e. $F=0$.
By direct inspection we find that this embedding satisfies the
conditions\footnote{In the STF case of Table 1.}
 (\ref{susyeq1}) and (\ref{susyeq2}), and hence is supersymmetric.
It preserves the translational invariance of $\varphi_1$ and $\varphi_2$. We then
expect the $U(1)_a$ and $U(1)_b$ symmetries to be preserved in the dual gauge 
theory description. 

This embedding and the dual flavoured gauge theory can be explained as follows. 
We have three sets of $N_f$
$D7$ branes located at
$z_1=m$, $z_2=m$, $z_3=m$ respectively, each one
supporting a flavour group $SU(N_f)$. We can join these 
branes \`a la Karch and Katz \cite{KarchKatz} 
and obtain one single set of $N_f$ $D7$ branes located as in (\ref{embd7lm}).
These $D7$-branes terminate before reaching the IR region and the
conformal invariance is explicitly broken by the mass $m$, which also
breaks the flavour groups $SU(N_f) \times SU(N_f) \times SU(N_f)$  
to the diagonal subgroup.

In order to deduce the superpotential of the dual gauge theory we
observe that
the same configuration can be 
realized
in the undeformed
($\gamma=0$, $AdS_5 \times S^5$ ) case;
here the superpotential is the 
following\footnote{We set the couplings to one for simplicity.}
\be
\label{supflav}
W=W_{N=4}+\tr Q_1 \Phi_1 \tilde Q_1+\tr Q_2 \Phi_2 \tilde Q_2+\tr Q_3 \Phi_3 \tilde Q_3+
m \, \tr (Q_1 \tilde Q_2+Q_2 \tilde Q_3+Q_3 \tilde Q_1 )
\ee 
Note that the massive flavours preserves the
$U(1)_a \times U(1)_b$ symmetry, 
assigning the charges as in Table 2. 
\begin{center}
\begin{tabular}{c|c|c|c|c|c|c|c|c|c}
&$\Phi_1$&$\Phi_2$&$\Phi_3$&$Q_1$&$\tilde Q_1$&$Q_2$&$\tilde Q_2$&$Q_3$&$\tilde Q_3$\\
\hline
$U(1)_a$&0&1&-1&1&-1&0&-1&1&0\\
\hline
$U(1)_b$&-1&1&0&0&1&-1&0&-1&1\\
\end{tabular}\\
Table 2
\end{center}
Now, for $N_f$ $D7$ branes embedded as (\ref{embd7lm}) in the beta deformed background,
the dual gauge theory is beta deformed
$\mathcal{N}=1$ SYM 
coupled to three different massive flavours. The resulting
 phase factors of the terms in the superpotential (\ref{supflav})
can be easily obtained following the prescription of \cite{LuninMalda} with the charges
in Table 2, having
\bea
W&=&
W_{\beta=\gamma}+
   e^{- i  \pi \gamma} \, \tr Q_1 \Phi_1 \tilde Q_1+ e^{ i  \pi \gamma}\, \tr Q_2 \Phi_2 \tilde Q_2+
  e^{- i  \pi \gamma} \, \tr Q_3 \Phi_3 \tilde Q_3+\non\\
  \label{supfladef}
&& + m \,  \tr (Q_1 \tilde Q_2+Q_2 \tilde Q_3+Q_3 \tilde Q_1 )  
\eea
Note that the flavour mass terms are not affected by the beta deformation.

\paragraph{Other $D7$ embeddings}
If we do not require the $U(1)_a$ and $U(1)_b$ global symmetries
to be preserved we can try to embed the $D7$ branes in other 
submanifolds, with vanishing world volume flux. 
The computations of the supersymmetry conditions (\ref{susyeq1}) and
(\ref{susyeq2}) get more complicated.

We take the embeddings
\bea
&&\xi_{\mu}=x_{\mu} \qquad \mu=0,\dots,3 \non \\
&& z_i=\xi_{4} e^{i \xi_{6}} \qquad z_j=\xi_{5} e^{i \xi_{7}} \qquad  z_k=m_0 
\qquad \qquad  i \neq j \neq k =1,2,3
\eea
A long computation shows they are supersymmetric for any choice of the mass $m_0$.
Here the dual gauge theory is beta deformed $\mathcal{N}=1$ SYM
plus $N_f$ flavours\footnote{$N_f$ is the number of $D7$ branes.}
which couple with the adjoint field $\Phi_k$.

Finally,
after a long computation, we find that the following 
$D7$ embeddings with chiral symmetry breaking
are supersymmetric
\bea
&&\xi_{\mu}=x_{\mu} \qquad \mu=0,\dots,3 \non \\
&& z_i=\xi_{4} e^{i \xi_{6}} \qquad z_j=\xi_{5} e^{i \xi_{7}} \qquad  z_k=\frac{m_0^2}{\xi_{5} e^{i \xi_{7}}} 
\qquad \qquad  i \neq j \neq k =1,2,3
\eea
The dual gauge theory is beta deformed $\mathcal{N}=1$ SYM with two kinds of $N_f$
massive flavours $Q_1$ and $Q_2$, which couple
to $\Phi_j$ and $\Phi_k$, respectively. The mass $m_0$ breaks the flavour
groups $SU(N_f)_1 \times SU(N_f)_2$ to the diagonal subgroup.

For these additional $D7$ embeddings the superpotential terms and their phase factors
 can be obtained 
  with the same procedure followed 
  in the derivation of (\ref{supfladef}), by starting
  from the $\mathcal{N}=4$ case
(i.e. $\gamma=0$).

\subsection{Effective Strings}
Finally we take $D$-branes that fill just two coordinates in the Minkowski space time
($\xi_0=x_0,\xi_1=x_1$). We place them at $x_2=x_3=0$.
We do not find supersymmetric configurations of $D3$ or $D5$ branes. 
We instead find that a
$D7$-brane 
 covering the whole internal space
\be
z_k=\xi_{k+1}+i \xi_{k+4} \qquad k=1,\dots,3
\ee
is supersymmetric.

\acknowledgments
I would like to 
thank especially L. Girardello, L. Martucci and A. Zaffaroni for 
useful 
discussions. 
I would also like to thank 
A. Amariti, A. Butti, R. Casero, D. Forcella, L. Mazzanti, S. McReynolds 
and M. Pirrone
for conversations. 
This work has been supported in part by INFN, PRIN prot.2005024045-002 and the
European Commission RTN program MRTN-CT-2004-005104.\\

\appendix
\section{The supersymmetric family of backgrounds and IR PW}
The supersymmetry equations for the ansatz (\ref{ansatz1},\ref{ansatz2}) 
was studied in 
\cite{Zaffa}. 
They imply, for complex solutions with constant dilaton, that 
the geometrical
quantities can be expressed as derivatives of a single function $F$. 
If the background
does not depend on $\sigma_3$ we have
\bea
\label{ai1}
&& A_{i\bar j} =  \frac{\partial^2 F}{\partial z_i \partial \bar{z}_j }  
\quad i,j=1,2 \,  \\
\label{ai2}
&& A_{i\bar j} \bar{\alpha}_j =   \frac{\partial^2 F}{\partial z_i \partial \bar{z}_3 }  \, ,\\
\label{ai3}
&& \alpha_i A_{i \bar j}  =    \frac{\partial^2 F}{\partial 
\bar{z}_j \partial z_3 } \, ,\\
\label{ai4}
&& u^2 a_3 \cos 2\phi + \alpha_i A_{i\bar j} \bar{\alpha}_j =   
\frac{\partial^2 F}{\partial z_3 \partial \bar{z}_3 } \, .\\
\label{fl} 
&& a_3 u^2 \sin^2 \phi = -  \frac{\partial}{\partial z_3} F \, .
\eea
The infrared geometry of the PW flow can be reconstructed in this
family of supersymmetric backgrounds as follows \cite{Zaffa}.
Choose coordinates
\bea
&& e^{z_1}=r^{3/4} \cos\theta \cos \varphi e^{i\sigma_1} \, ,\nonumber\\
&& e^{z_2}=r^{3/4} \cos\theta \sin \varphi e^{i\sigma_2} \, ,\nonumber\\
&& e^{z_3}=r^{3/2} \sin\theta e^{i\sigma_3} \, \nonumber .
\eea
The generalized Kahler potential $F$ is
\beq 
F = \frac{3}{4} r^2 (1- 2 \sin^2 \theta ) \, ,
\eeq
and the warp factor 
\be
e^{2A}=r^2 \sqrt{\frac{3}{4} (1+\sin^2 \theta)}
\ee
The other quantities are determined, for example
\bea
\sin 2 \phi&=&\frac{\sin \theta \sqrt{2+\sin^2 \theta}}{1+\sin^2 \theta} \\
A_{1 \bar 1}&=& r^2  \left(\cos^2 \theta \cos^2 \varphi+
\frac{\cos^4 \theta \cos^4 \varphi}{3 + 3\sin^2 \theta}  \right)\\
A_{1 \bar 2}=A_{2 \bar 1}&=& \frac{r^2 \cos^4 \theta \sin^2 \varphi \cos^2 \varphi}{3+3 \sin^2 \theta}\\
A_{2 \bar 2}&=& r^2  \left(\cos^2 \theta \sin^2 \varphi+
\frac{\cos^4 \theta \sin^4 \varphi}{3 + 3\sin^2 \theta}  \right) \\
a_3&=&\frac{1+\sin^2 \theta}{4 r (2+\sin^2 \theta)}
\eea

\section{Beta deformed gravity dual}
We have already introduced the complex coordinates $z_i$ (\ref{complexLM});
 the one forms appearing in 
(\ref{metricazaf}) are defined as \cite{Zaffa}
\bea
&& x_1+i y_1=e^{- i \sigma_1} \sqrt{\frac{g}{\mu_1^2(\mu_2^2+\mu_3^2)}} (d z_1-\frac{\bar z_2 \bar z_3 z}{r^2 \sqrt{g}}) \\
&& x_2+i y_2=e^{-i \sigma_2} \sqrt{1+\frac{\mu_3^2}{\mu_2^2}}(d z_2-\frac{\bar z_1 \bar z_3 z}{r^2 \sqrt{g}})+\frac{\mu_3^2 e^{-i \sigma_1}}{\mu_1 \sqrt{\mu_2^2+\mu_3^2}}(dz_1-\frac{\bar z_2 \bar z_3 z}{r^2 \sqrt{g}}) \\
&&z=\frac{d[z_1 z_2 z_3]}{r^2 \sqrt{g}}
\eea
The internal metric (\ref{metricazaf}) gives then \cite{LuninMalda}
\be
 d \tilde s^2_6=dr^2+r^2 \left( \sum_{i=1}^3(d \mu_i^2+G \mu_i^2 d \sigma_i^2 )+
\gamma^2 G \mu_1^2 \mu_2^2 \mu_3^2 (d\sigma_1+d\sigma_2+d\sigma_3)^2 \right)
\ee


\begin{thebibliography}{99}

\bibitem{MaldaWitten}
J.~M.~Maldacena,
  ``The large N limit of superconformal field theories and supergravity,''
  Adv.\ Theor.\ Math.\ Phys.\  {\bf 2} (1998) 231
  [Int.\ J.\ Theor.\ Phys.\  {\bf 38} (1999) 1113]
  [arXiv:hep-th/9711200].

S.~S.~Gubser, I.~R.~Klebanov and A.~M.~Polyakov,
  ``Gauge theory correlators from non-critical string theory,''
  Phys.\ Lett.\  B {\bf 428} (1998) 105
  [arXiv:hep-th/9802109].

  E.~Witten,
  ``Anti-de Sitter space and holography,''
  Adv.\ Theor.\ Math.\ Phys.\  {\bf 2} (1998) 253
  [arXiv:hep-th/9802150].

O.~Aharony, S.~S.~Gubser, J.~M.~Maldacena, H.~Ooguri and Y.~Oz,
  ``Large N field theories, string theory and gravity,''
  Phys.\ Rept.\  {\bf 323} (2000) 183
  [arXiv:hep-th/9905111].


\bibitem{Mariana}
M.~Grana,
  ``Flux compactifications in string theory: A comprehensive review,''
  Phys.\ Rept.\  {\bf 423} (2006) 91
  [arXiv:hep-th/0509003].


\bibitem{GKP}
S.~B.~Giddings, S.~Kachru and J.~Polchinski,
  ``Hierarchies from fluxes in string compactifications,''
  Phys.\ Rev.\  D {\bf 66} (2002) 106006
  [arXiv:hep-th/0105097].


\bibitem{Grana1}
M.~Grana, R.~Minasian, M.~Petrini and A.~Tomasiello,
  ``Supersymmetric backgrounds from generalized Calabi-Yau manifolds,''
  JHEP {\bf 0408} (2004) 046
  [arXiv:hep-th/0406137].

 M.~Grana, R.~Minasian, M.~Petrini and A.~Tomasiello,
  ``Type II strings and generalized Calabi-Yau manifolds,''
  Comptes Rendus Physique {\bf 5} (2004) 979
  [arXiv:hep-th/0409176].
 
\bibitem{Grana2}

M.~Grana, R.~Minasian, M.~Petrini and A.~Tomasiello,
  ``Generalized structures of N=1 vacua,''
  JHEP {\bf 0511} (2005) 020
  [arXiv:hep-th/0505212].



\bibitem{Gauntlett}
  J.~P.~Gauntlett, D.~Martelli, S.~Pakis and D.~Waldram,
  ``G-structures and wrapped NS5-branes,''
  Commun.\ Math.\ Phys.\  {\bf 247} (2004) 421
  [arXiv:hep-th/0205050].

J.~P.~Gauntlett, D.~Martelli, J.~Sparks and D.~Waldram,
  ``Supersymmetric AdS(5) solutions of type IIB supergravity,''
  Class.\ Quant.\ Grav.\  {\bf 23} (2006) 4693
  [arXiv:hep-th/0510125].


    
\bibitem{Hitchin}
N.~Hitchin,
  ``Generalized Calabi-Yau manifolds,''
  Quart.\ J.\ Math.\ Oxford Ser.\  {\bf 54} (2003) 281
  [arXiv:math.dg/0209099].


\bibitem{Gualtieri}
M.~Gualtieri,
  ``Generalized complex geometry,''
  arXiv:math.dg/0401221.
Oxford University DPhil thesis, arXiv:math.DG/0401221.






\bibitem{esseu3}
I.~R.~Klebanov and M.~J.~Strassler,
  ``Supergravity and a confining gauge theory: Duality cascades and
  chiSB-resolution of naked singularities,''
  JHEP {\bf 0008} (2000) 052
  [arXiv:hep-th/0007191].
  
  J.~M.~Maldacena and C.~Nunez,
  ``Towards the large N limit of pure N = 1 super Yang Mills,''
  Phys.\ Rev.\ Lett.\  {\bf 86} (2001) 588
  [arXiv:hep-th/0008001].


A.~Butti, M.~Grana, R.~Minasian, M.~Petrini and A.~Zaffaroni,
  ``The baryonic branch of Klebanov-Strassler solution: A supersymmetric
  family of SU(3) structure backgrounds,''
  JHEP {\bf 0503} (2005) 069
  [arXiv:hep-th/0412187].


\bibitem{Dallagata}
G.~Dall'Agata,
  ``On supersymmetric solutions of type IIB supergravity with general
  fluxes,''
  Nucl.\ Phys.\  B {\bf 695} (2004) 243
  [arXiv:hep-th/0403220].





\bibitem{Zaffa}
R.~Minasian, M.~Petrini and A.~Zaffaroni,
  ``Gravity duals to deformed SYM theories and generalized complex geometry,''
  JHEP {\bf 0612} (2006) 055
  [arXiv:hep-th/0606257].



\bibitem{PW}
K.~Pilch and N.~P.~Warner,
  ``A new supersymmetric compactification of chiral IIB supergravity,''
  Phys.\ Lett.\  B {\bf 487} (2000) 22
  [arXiv:hep-th/0002192].

K.~Pilch and N.~P.~Warner,
  ``N = 1 supersymmetric renormalization group flows from IIB supergravity,''
  Adv.\ Theor.\ Math.\ Phys.\  {\bf 4} (2002) 627
  [arXiv:hep-th/0006066].

K.~Pilch and N.~P.~Warner,
  ``N = 1 supersymmetric solutions of IIB supergravity from Killing  spinors,''
  arXiv:hep-th/0403005.


C.~N.~Gowdigere and N.~P.~Warner,
  ``Holographic Coulomb branch flows with N = 1 supersymmetry,''
  JHEP {\bf 0603} (2006) 049
  [arXiv:hep-th/0505019].


\bibitem{LuninMalda}
O.~Lunin and J.~M.~Maldacena,
  ``Deforming field theories with U(1) x U(1) global symmetry and their
  gravity duals,''
  JHEP {\bf 0505} (2005) 033
  [arXiv:hep-th/0502086].



\bibitem{Koerber}
  P.~Koerber,
  ``Stable D-branes, calibrations and generalized Calabi-Yau geometry,''
  JHEP {\bf 0508} (2005) 099
  [arXiv:hep-th/0506154].

\bibitem{Martucci}
L.~Martucci and P.~Smyth,
  ``Supersymmetric D-branes and calibrations on general N = 1 backgrounds,''
  JHEP {\bf 0511} (2005) 048
  [arXiv:hep-th/0507099].

\bibitem{Martucci2}
L.~Martucci,
  ``D-branes on general N = 1 backgrounds: Superpotentials and D-terms,''
  JHEP {\bf 0606} (2006) 033
  [arXiv:hep-th/0602129].

  L.~Martucci,
  ``Supersymmetric D-branes on flux backgrounds,''
  arXiv:hep-th/0701093.

P.~Koerber and L.~Martucci,
  ``Deformations of calibrated D-branes in flux generalized complex
  manifolds,''
  JHEP {\bf 0612} (2006) 062
  [arXiv:hep-th/0610044].




\bibitem{Ooguri}
O.~DeWolfe, D.~Z.~Freedman and H.~Ooguri,
 ``Holography and defect conformal field theories,''
  Phys.\ Rev.\  D {\bf 66} (2002) 025009
  [arXiv:hep-th/0111135].


\bibitem{KarchKatz}
 A.~Karch and E.~Katz,
 ``Adding flavor to AdS/CFT,''
  JHEP {\bf 0206} (2002) 043
  [arXiv:hep-th/0205236].


\bibitem{Nonabe} 
D.~Berenstein and R.~G.~Leigh,
  ``Discrete torsion, AdS/CFT and duality,''
  JHEP {\bf 0001} (2000) 038
  [arXiv:hep-th/0001055].


\bibitem{Mika}
A.~Mikhailov,
  ``Giant gravitons from holomorphic surfaces,''
  JHEP {\bf 0011} (2000) 027
  [arXiv:hep-th/0010206].

\bibitem{Marco}
M.~Pirrone,
  ``Giants on deformed backgrounds,''
  JHEP {\bf 0612} (2006) 064
  [arXiv:hep-th/0609173].


\bibitem{LS}
R.~G.~Leigh and M.~J.~Strassler,
  ``Exactly Marginal Operators And Duality In Four-Dimensional N=1
  Supersymmetric Gauge Theory,''
  Nucl.\ Phys.\  B {\bf 447} (1995) 95
  [arXiv:hep-th/9503121].


\bibitem{Randall}
A.~Karch and L.~Randall,
  ``Open and closed string interpretation of SUSY CFT's on branes with
  boundaries,''
  JHEP {\bf 0106} (2001) 063
  [arXiv:hep-th/0105132].



\bibitem{BF}
P.~Breitenlohner and D.~Z.~Freedman,
  ``Stability In Gauged Extended Supergravity,''
  Annals Phys.\  {\bf 144} (1982) 249.

P.~Breitenlohner and D.~Z.~Freedman,
  ``Positive Energy In Anti-De Sitter Backgrounds And Gauged Extended
  Supergravity,''
  Phys.\ Lett.\  B {\bf 115} (1982) 197.







\end{thebibliography}
\end{document}